\documentclass[twocolumn,prb,showpacs,preprintnumbers,amsmath,amssymb,superscriptaddress]{revtex4}

\usepackage{subfigure}
\usepackage{graphicx}
\usepackage{dcolumn}
\usepackage{bm}
\usepackage{epsfig}

\begin{document}


\title{Antiferromagnetic order and domains in $\mathrm{Sr_{3}Ir_{2}O_{7}}$ probed by x-ray resonant scattering}

\author{S. Boseggia}
\email{stefano.boseggia@diamond.ac.uk}%
\affiliation{London Centre for Nanotechnology and Department of Physics and Astronomy, University College London, London WC1E 6BT, United Kingdom}
\affiliation{Diamond Light Source Ltd., Oxfordshire OX11 0DE, United Kingdom}
\author{R. Springell}
\affiliation{London Centre for Nanotechnology and Department of Physics and Astronomy, University College London, London WC1E 6BT, United Kingdom}
\author{H. C. Walker}
\affiliation{European Synchrotron Radiation Facility, BP220, F-38043 Grenoble Cedex 9, France}
\author{A. T. Boothroyd}
\author{D. Prabhakaran}
\affiliation{Clarendon Laboratory, Department of Physics, University of Oxford, Parks Road, Oxford OX1 3PU, UK}
\author{D. Wermeille}
\author{L. Bouchenoire}
\affiliation{XMaS, UK-CRG, European Synchrotron Radiation Facility, BP220, F-38043 Grenoble Cedex, France}
\affiliation{Oliver Lodge Laboratory, Department of Physics, University of Liverpool, Oxford Street, Liverpool, L69 7ZE, UK}
\author{S. P. Collins}
\affiliation{Diamond Light Source Ltd., Oxfordshire OX11 0DE, United Kingdom}
\author{D. F. McMorrow}
\affiliation{London Centre for Nanotechnology and Department of Physics and Astronomy, University College London, London WC1E 6BT, United Kingdom}
\date{\today}

\begin{abstract}
This article reports a detailed x-ray resonant scattering study of the bilayer iridate compound, $\mathrm{Sr_{3}Ir_{2}O_{7}}$, at the Ir $\mathrm{L_{2}}$ and $\mathrm{L_{3}}$ edges. Resonant scattering at the Ir $\mathrm{L_{3}}$ edge has been used to determine that $\mathrm{Sr_{3}Ir_{2}O_{7}}$ is a long-range ordered antiferromagnet below $\mathrm{T_{N}\approx230\,K}$ with an ordering wavevector, $\mathrm{\mathbf{q}=(\frac{1}{2},\frac{1}{2},0)}$. The energy resonance at the $\mathrm{L_{3}}$ edge was found to be a factor of $\sim30$ times larger than that at the $\mathrm{L_{2}}$. This remarkable effect has been seen in the single layer compound $\mathrm{Sr_{2}IrO_{4}}$ and has been linked to the observation of a $\mathrm{J_{eff}=\frac{1}{2}}$ spin-orbit insulator. Our result shows that despite the modified electronic structure of the bilayer compound, caused by the larger bandwidth, the effect of strong spin-orbit coupling on the resonant magnetic scattering persists. Using the programme SARA\textit{h}, we have determined that the magnetic order consists of two domains with propagation vectors $\mathrm{\mathbf{k_{1}}=(\frac{1}{2},\frac{1}{2},0)}$ and $\mathrm{\mathbf{k_{2}}=(\frac{1}{2},-\frac{1}{2},0)}$, respectively. A raster measurement of a focussed x-ray beam across the surface of the sample yielded images of domains of the order of $\mathrm{100\,\mu m}$ size, with odd and even \textit{l} components, respectively. Fully relativistic, monoelectronic calculations (FDMNES), using the Green's function technique for a muffin-tin potential have been employed to calculate the relative intensities of the $\mathrm{L_{2,3}}$ edge resonances, comparing the effects of including spin-orbit coupling and the Hubbard, \textit{U}, term. A large $\mathrm{L_{3}}$ to $\mathrm{L_{2}}$ edge intensity ratio ($\sim5$) was found for calculations including spin-orbit coupling. Adding the Hubbard, U, term resulted in changes to the intensity ratio $\mathrm{<5\,\%}$.
\end{abstract}

\pacs{numbers}
\maketitle

\section{\label{intro}Introduction}

Transition metal oxides (TMOs) are responsible for an enormous range of phenomena in solid state physics \cite{Cheong}, largely dominated by materials containing ions of the \textit{3d} series. Studies into these compounds have not only enriched our fundamental understanding of materials, but have resulted in the realization of applications in the real world; from high temperature superconductivity in the cuprates \cite{Bednorz} to colossal magnetoresistance (CMR) in perovskite manganites \cite{Ramirez} and more recently, multiferroicity \cite{Cheong2}. It was inevitable that further advances would be made and new physics would be discovered in the heavier \textit{4d} and \textit{5d} series. As the transition metal ion species becomes heavier, relativistic effects start to play a more important role and the competing energy scales of the crystal field and spin-orbit coupling present a host of new possibilities \cite{Kim2,Shitade,Jiang,Machida}.

Conventional wisdom concerning TMOs describes the \textit{3d} compounds as having a localized, strongly correlated, narrow \textit{d} band with a large value for the Coulomb repulsion, \textit{U}. \textit{5d} TMOs by contrast were thought to have weakly-correlated, wide \textit{d} bands with a reduced Coulomb term. One might then expect the \textit{5d} TMOs to have a greater tendency to be metallic. However, in a number of notable cases, for example $\mathrm{Sr_{2}IrO_{4}}$ \cite{Crawford}, $\mathrm{Sr_{3}Ir_{2}O_{7}}$ \cite{Cao} and $\mathrm{Na_{2}IrO_{3}}$ \cite{Singh}, insulating or poor metallic behavior has been observed. The traditional origin of the Mott insulator, namely that \textit{U} is similar in magnitude to the band width, \textit{W}, is not applicable in these cases, since \textit{U} is greatly reduced. The most likely culprit is then the spin-orbit (SO) coupling, and in fact recently, the first example of a spin-orbit Mott insulator was proposed to explain the experimental data observed from the layered iridate compound, $\mathrm{Sr_{2}IrO_{4}}$ \cite{Kim}.

Supported by density functional theory calculations with local density approximation (LDA) $+\,U\,+\,\mathrm{SO}$, angular resolved photoemission spectroscopy (ARPES), x-ray absorption (XAS), optical conductivity data \cite{Kim} and later, x-ray resonant scattering (XRS) \cite{Kim2}, a convincing argument was made for the role of the SO energy $\zeta_{SO}$, in the observed insulating behavior and a novel, $\mathrm{J_{eff}=\frac{1}{2}}$ Mott ground state. The Ir \textit{5d} levels are split by the crystal field (CF) into $t_{\mathrm{2g}}$ and $e_{\mathrm{g}}$ orbitals. The CF energies are large which results in a low spin state, $t_{\mathrm{2g}}^{5}$. In the limit of strong SO coupling the $t_{\mathrm{2g}}$ band is split into a $\mathrm{J_{eff}=\frac{1}{2}}$ doublet and a $\mathrm{J_{eff}=\frac{3}{2}}$ quartet, which is lower in energy. The $\mathrm{J_{eff}=\frac{3}{2}}$ band is filled by four of the Ir $5d^{5}$ electrons, leaving a half-filled $\mathrm{J_{eff}=\frac{1}{2}}$ band. The Coulomb repulsion, \textit{U}, splits the $\mathrm{J_{eff}=\frac{1}{2}}$ band, giving rise to a Mott insulator; the first example of a `spin-orbit integrated narrow band system' \cite{Kim}.

$\mathrm{Sr_{3}Ir_{2}O_{7}}$ is the next compound along the Ruddlesden Popper series, $\mathrm{Sr_{n+1}Ir_{n}O_{3n+1}}$ from $\mathrm{Sr_{2}IrO_{4}}$. It can be described as a bilayered iridate, which displays bulk properties \cite{Cao,Nagai} with some similarities to its single layered cousin \cite{Cao2,Kini}. It is a weakly insulating magnetic compound with a tetragonal I4/\textit{mmm} crystal structure at room temperature \cite{Cao}. A weak ferromagnetic component is observed to appear in the magnetization data at $\mathrm{T_{A}\sim280\,K}$ and a second transition, resulting in a further increase in magnetization occurs at $\mathrm{T_{B}\sim250\,K}$ \cite{Cao}, attributed to spin-canting of an antiferromagnetically ordered state. The magnetic moment reported at $\mathrm{2\,K}$ in an applied field of $\mathrm{7\,T}$ is only $\mathrm{0.037\,\mu_{B}/Ir}$, more than 25 times lower than the expected $\mathrm{1\,\mu_{B}/Ir}$ for an $\mathrm{S=\frac{1}{2}}$ system. The effective moment, deduced from a high-temperature Curie-Weiss (CW) fit to the magnetic susceptibility, $\chi$, in a field of $\mathrm{7\,T}$ was reported to be $\mathrm{0.69\mu_{B}/Ir}$, compared to the theoretical value of $\mathrm{1.73\mu_{B}/Ir}$ for an $\mathrm{S=\frac{1}{2}}$ system. These results are consistent with a picture of a canted antiferromagnetic, $\mathrm{J_{eff}=\frac{1}{2}}$ insulator, similar to the single layered $\mathrm{Sr_{2}IrO_{4}}$ compound. Despite these similarities, $\mathrm{Sr_{3}Ir_{2}O_{7}}$ is much more metallic than $\mathrm{Sr_{2}IrO_{4}}$, reflecting the general tendency towards greater metallicity with increasing n in the $\mathrm{Sr_{n+1}Ir_{n}O_{3n+1}}$ series \cite{Moon}. A question then remains as to the robustness of the properties driven by the strong spin-orbit coupling, as the electron correlation strength is changed.

In order to investigate the nature of the antiferromagnetic ordering of the Ir magnetic moments, the technique of neutron diffraction would normally be the first port of call. However, due to the strong absorption cross-section for thermal neutrons of both naturally occurring iridium isotopes, $\mathrm{^{191}Ir}$ $\mathrm{(954\,barn)}$ $\mathrm{^{193}Ir}$ $\mathrm{(111\,barn)}$ and the small magnetic moment, x-ray resonant scattering (XRS) is the most suitable tool to investigate the magnetic structure of $\mathrm{Sr_{3}Ir_{2}O_{7}}$. It is both an element selective and shell specific technique and can be used to measure systems with small magnetic moments \cite{Isaacs}. The origin of the magnetic scattering is due to electric multipole transitions from core levels to spin-polarized, empty states in the valence band. The intensity of the scattering is determined by the SO splitting of the initial and intermediate states and the overlap integral of the intermediate levels with the initial and final electronic orbitals. For the case of $\mathrm{Sr_{3}Ir_{2}O_{7}}$ this effect is large at the Ir $\mathrm{L_{2,3}}$, which probe dipole transitions from $2p_{1/2}$, $2p_{3/2}$ core levels, respectively, to $5d_{3/2,5/2}$ states.

Recent work by \textit{Kim et al.} \cite{Kim2} proposed that there may be an even greater potential behind the XRS technique, interpreting the relative strength of the magnetic scattering at the $\mathrm{L_{2}}$ and $\mathrm{L_{3}}$ edges to suggest a possible $\mathrm{J_{eff}=\frac{1}{2}}$ ground state. However, the interpretation of these data has been disputed \cite{Chapon}. A major question remains as to the robustness of the potential $\mathrm{J_{eff}=\frac{1}{2}}$ state for such dramatic changes in the electronic properties. This article presents a detailed account of a resonant x-ray scattering investigation into the barely insulating bilayered iridate, $\mathrm{Sr_{3}Ir_{2}O_{7}}$. The aim is to determine the magnetic structure of this compound and to investigate the unusual behavior of the $\mathrm{L_{2,3}}$ edge resonances, attributed to an effect of the $\mathrm{J_{eff}=\frac{1}{2}}$ state \cite{Kim2}, and to determine if this effect persists in a larger bandwidth, less insulating system.

The article is organized as follows: Section \ref{structure} describes the crystal structure determination and section \ref{magnetization}, the bulk susceptibility data, comparing with previous reports. Section \ref{xrs} is an x-ray resonant scattering (XRS) investigation of the magnetic order, which includes a study of the energy, polarization and temperature dependences of the resonant magnetic peaks, a Group Theory treatment of the magnetic structure and a domain imaging measurement. Section \ref{FDMNES} reports calculations of the energy resonances and compares with experimental data. We conclude that the magnetic structure is a two-domain commensurate antiferromagnet with propagation vectors, $\mathrm{\mathbf{k_{1}}=(\frac{1}{2},\frac{1}{2},0)}$ and $\mathrm{\mathbf{k_{2}}=(\frac{1}{2},-\frac{1}{2},0)}$ and a N\'{e}el temperature, $\mathrm{T_{N}=230\pm5\,K}$. The exceptionally large ratio of Ir $\mathrm{L_{3}}$ to $\mathrm{L_{2}}$ edge magnetic scattering intensity that was seen in $\mathrm{Sr_{2}IrO_{4}}$ \cite{Kim2} is also present in this compound, suggesting that the $\mathrm{J_{eff}=\frac{1}{2}}$ state may still be realized in this much wider bandwidth compound.

\section{Experimental results}

\subsection{\label{structure}Crystal structure determination}

\begin{figure}[h!]
\centering
\includegraphics[width=0.45\textwidth,bb=20 80 480 480,clip]{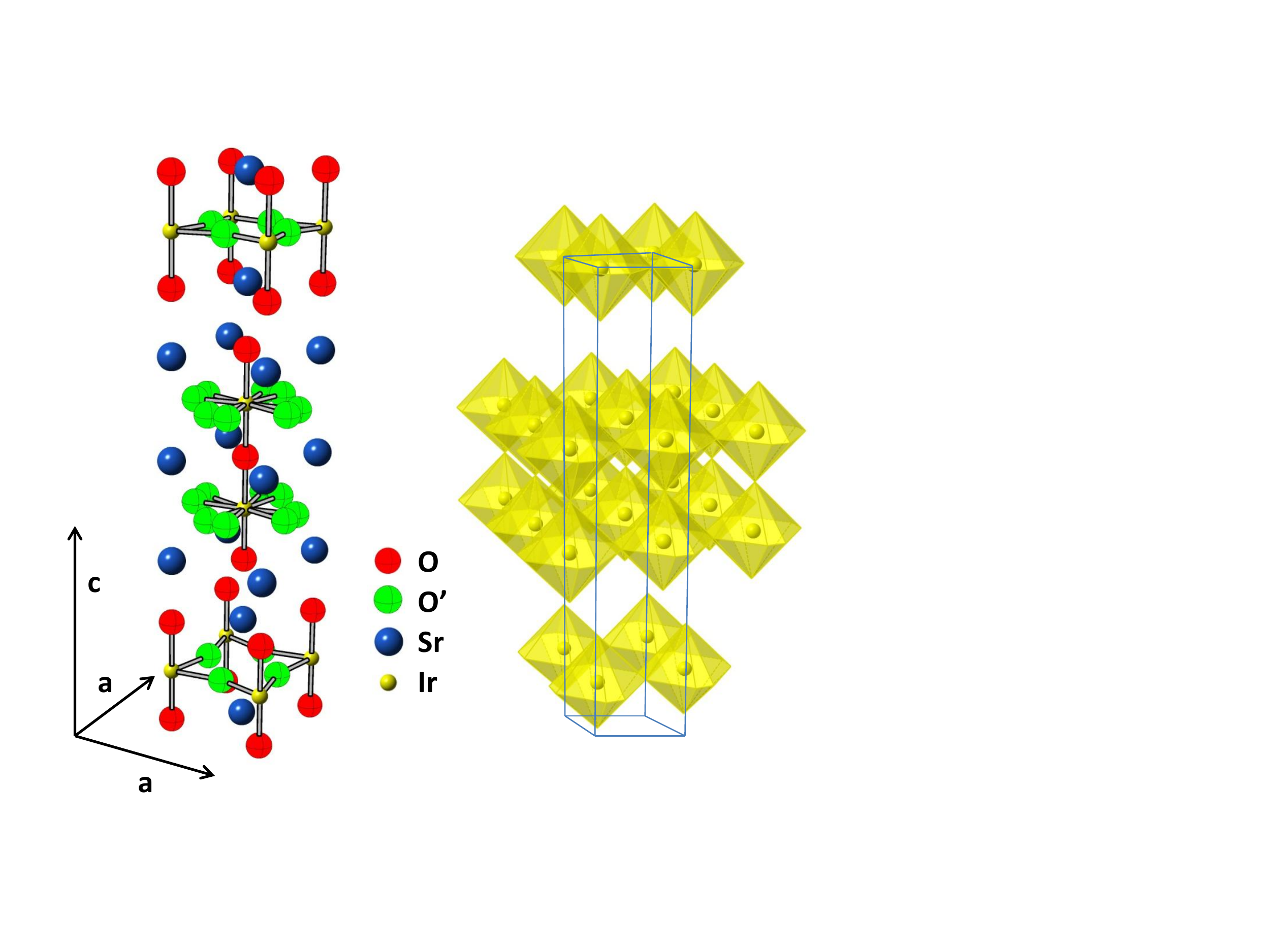}\caption
{\label{figure1}(Color online) The $\mathrm{Sr_{3}Ir_{2}O_{7}}$ I4/mmm crystal structure (left) with two bilayer molecules per unit cell. A 50\% occupancy of the O' sites results in an octahedral oxygen environment of the Ir ions, rotated from one layer to the next. A polyhedral representation of the Ir octahedral environment is shown on the right-hand-side.}
\end{figure}

Single crystals of $\mathrm{Sr_{3}Ir_{2}O_{7}}$ were grown at the Clarendon Laboratory, Oxford University, UK. These were synthesized in Pt crucibles using the self-flux technique from off-stoichiometric $\mathrm{IrO_{2}}$, $\mathrm{SrCO_{3}}$ and $\mathrm{SrCl_{2}}$ compounds. The mixture was heated to $\mathrm{1440^{\circ}\,C}$, fired for $20\,h$ and slowly cooled at $3^{\circ}/h$. The resulting samples were plate-like with the \textit{c}-axis along the shortest dimension of  $\mathrm{\sim2\times2\times0.1\,mm}$ size crystals.

$\mathrm{Sr_{3}Ir_{2}O_{7}}$ is the n=2 compound of the $\mathrm{Sr_{n+1}Ir_{n}O_{3n+1}}$ Ruddlesden-Popper series, whose synthesis was first reported in 1994 \cite{Subramanian}. However, the exact crystal structure remains a contentious issue. Previous articles have reported the tetragonal I4/\textit{mmm} \cite{Subramanian}, the orthorhombic Bbca \cite{Cao}, Bbcb \cite{Matsuhata} and Pban \cite{Radaelli} space groups. It is clear that the structure contains strongly coupled double Ir-O layers, separated by layers of Sr-O and off-set along the \textit{c}-axis, which result in a double-layered framework of Ir atoms centered inside oxygen octahedra. It is the rotation of these octahedra and the correlation between the rotations that lead to the subtle differences in crystal structure reported thus far.

X-ray diffraction data were collected with a Supernova x-ray diffractometer equipped with a microfocused monochromatic Mo source at the Research Complex at Harwell (RCaH), Chilton, UK.
A $\mathrm{Sr_{3}Ir_{2}O_{7}}$ single crystal of dimensions $\mathrm{0.108\times0.134\times0.072\,mm}$ was mounted. 1369 reflections (204 unique) were measured at room temperature in an $\omega$-scan mode within a $\mathrm{2\theta}$ range of $7.84^\circ$ to $64.6^\circ$. The data were corrected for Lorentz, polarization and absorption effects. The 1369 reflections were used to obtain the cell parameter: $a=3.897(5)$ and $c\mathrm{=20.892(5)\,\AA}$ and the calculated density was $\mathrm{7.947\, g/cm^{3}}$.

Initially, the diffraction data were modeled with an I4/\textit{mmm} space group and a full occupancy of the atom sites. As a consequence we obtained a large R factor ($\mathrm{>0.7}$) and the thermal parameter of O(3) (equivalent to O atomic positions, labelled O' in Fig. \ref{figure1} (a)) became very large compared to the other atoms and strongly anisotropic with an ellipse highly elongated along the \textit{a}-axis. A further step was then to remove the O(3) atom from its position on the mirror plane and shift it by $\mathrm{0.4\,\AA}$ assigning it an occupancy of 0.5. Subsequent refinement of several parameters (isotropic extinction, anisotropic thermal parameter for Ir and Sr, isotropic thermal parameters for O(3)) by full-matrix least-squares techniques converged at R=0.039. The crystal structure of $\mathrm{Sr_{3}Ir_{2}O_{7}}$ is shown in Fig. \ref{figure1} (left panel), the oxygen atoms highlighted in green represent those with only 50\% occupancy. This refinement of the lattice parameters and atomic positions is in close agreement with the initial report on the crystal structure of $\mathrm{Sr_{3}Ir_{2}O_{7}}$ by Subramanian \textit{et al.} \cite{Subramanian}.

The disorder of O atoms is associated with the alternate rotation of the $\mathrm{IrO_{6}}$ octahedra about the \textit{c}-axis. From our data this rotation is about $\mathrm{11.91^\circ}$. An octahedral representation of the structure is shown in Fig. \ref{figure1} (right panel). No additional superlattice reflections could be discerned in the data suggesting that the octahedral rotations are not correlated, neither in the plane nor perpendicular to it. Although it is possible that more subtle crystallographic effects might be elucidated using more intense x-ray sources, for the purpose of determining the magnetic structure, using the I4/\textit{mmm} space group is reasonable.

\subsection{\label{magnetization}Magnetization measurements}

\begin{figure}[h!]
\centering
\includegraphics[width=0.45\textwidth,bb=30 50 520 720,clip]{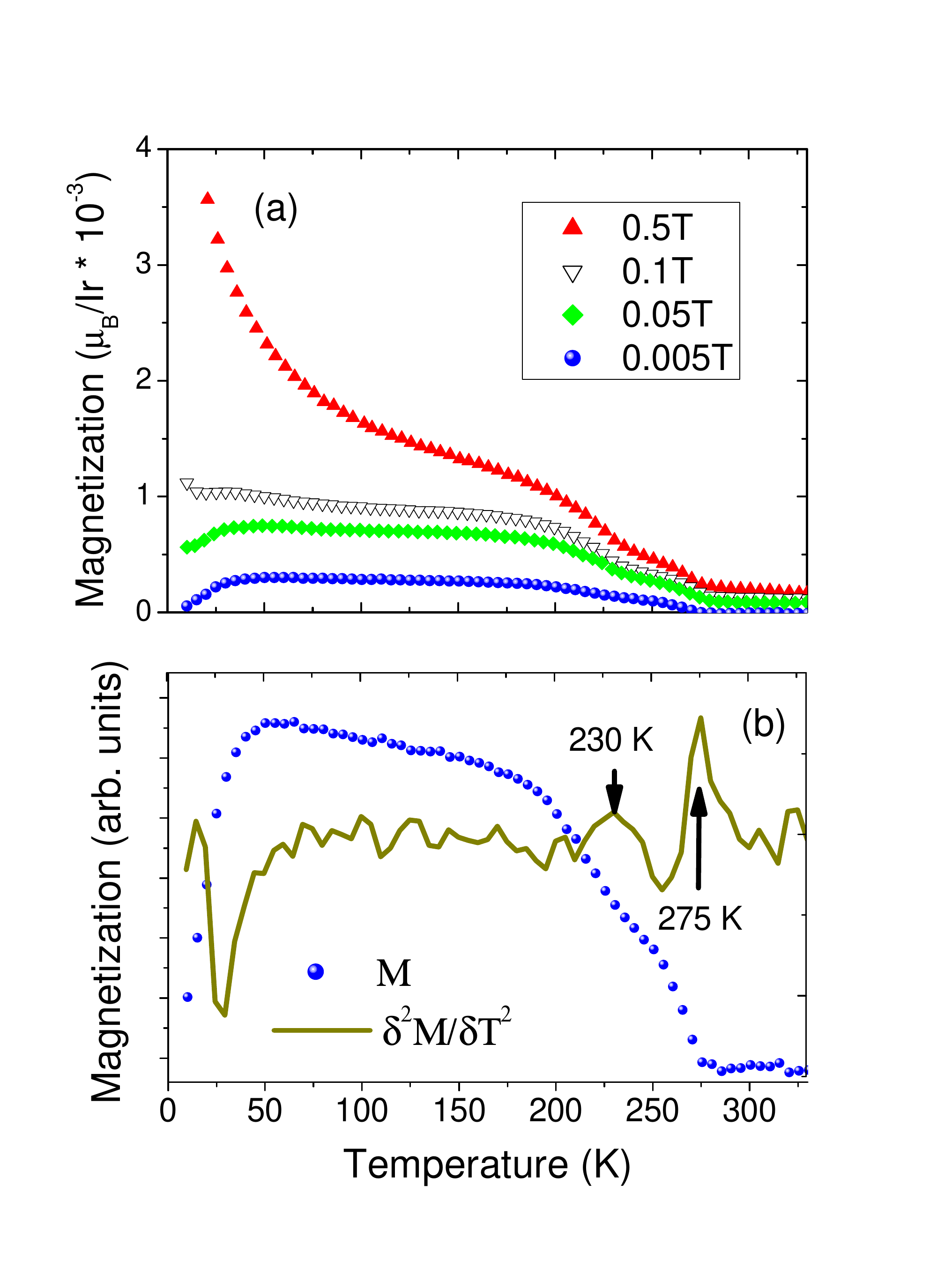}\caption
{\label{figure2}(Color online) SQUID magnetization data, as a function of temperature for field cooled (FC) measurements in the basal plane, ranging from 0.005\,T to 0.5\,T applied magnetic field, panel (a). Two transitions can be observed at high temperature and a downturn in the magnetization at 50\,K, which becomes a positive upturn as the field is increased. Panel (b) shows the 0.005\,T data, together with the second derivative of the magnetization; the peaks at 230\,K and 275\,K indicate the two transition temperatures.}
\end{figure}

The bulk magnetization data were collected, using a Quantum Design MPMS-7 superconducting quantum interference device (SQUID) magnetometer, at the London Centre for Nanotechnology, UCL, London, UK. Figure \ref{figure2} shows the magnetization, M ($\mathrm{\mu_{B}/Ir}$) as a function of temperature, T, measured in the basal plane. Field cooled (FC) measurements were made on cooling in an applied magnetic field, ranging from $\mathrm{0.05-0.5\,T}$. On inspection of the FC data, it is clear that there are two transitions at high temperature, $\mathrm{T_{A}}$ and $\mathrm{T_{B}}$, with a downturn in the magnetization at low temperature, beginning at approx. 50\,K. This downturn becomes an upturn in M at higher fields. Panel (b) of Fig. \ref{figure2} shows the second derivative of the 0.005\,T FC data, which gives $\mathrm{T_{A}=275\,K}$ and $\mathrm{T_{B}=230\,K}$.

These data are qualitatively similar to those reported previously on single crystals. A study by Cao \textit{et al.} reported transitions $\mathrm{T_{A}=285\,K}$, $\mathrm{T_{B}=260\,K}$ \cite{Cao2}, for FC magnetization measurements in the \textit{ab}-plane. The lower transition temperatures in our case may be due to variations in the sample quality or small differences in oxygen stoichiometry between the two studies.

\subsection{\label{xrs}X-ray resonant scattering}

The x-ray resonant scattering experiments were conducted at the I16 beamline of the Diamond Light Source, Didcot, UK and at the BM28 (XMaS) beamline \cite{Brown} at the European Synchrotron Radiation Facility (ESRF), Grenoble, France. Measurements were performed at the Ir $\mathrm{L_{2}}$ (12.831\,keV) and $\mathrm{L_{3}}$ (11.217\,keV) absorption edges, probing transitions from the $2p_{\frac{1}{2}}$ to the $5d_{\frac{3}{2}}$ and from the $2p_{\frac{3}{2}}$ to the $5d_{\frac{3}{2},\frac{5}{2}}$, states, respectively.

\subsubsection{Ordering wavevector}

On I16 the x-rays were provided by means of a U27-type undulator insertion device, focussed to a beam size of $20\times200\,\mu m$ at the sample position, using a set of parallel double focussing mirrors. A Newport 6-axis N-6050 Kappa diffractometer was used to maneuver the sample orientation and an avalanche photodiode (APD) was used to detect the scattered photons.

\begin{figure}[h!]
\centering
\includegraphics[width=0.475\textwidth,bb=15 30 530 740,clip]{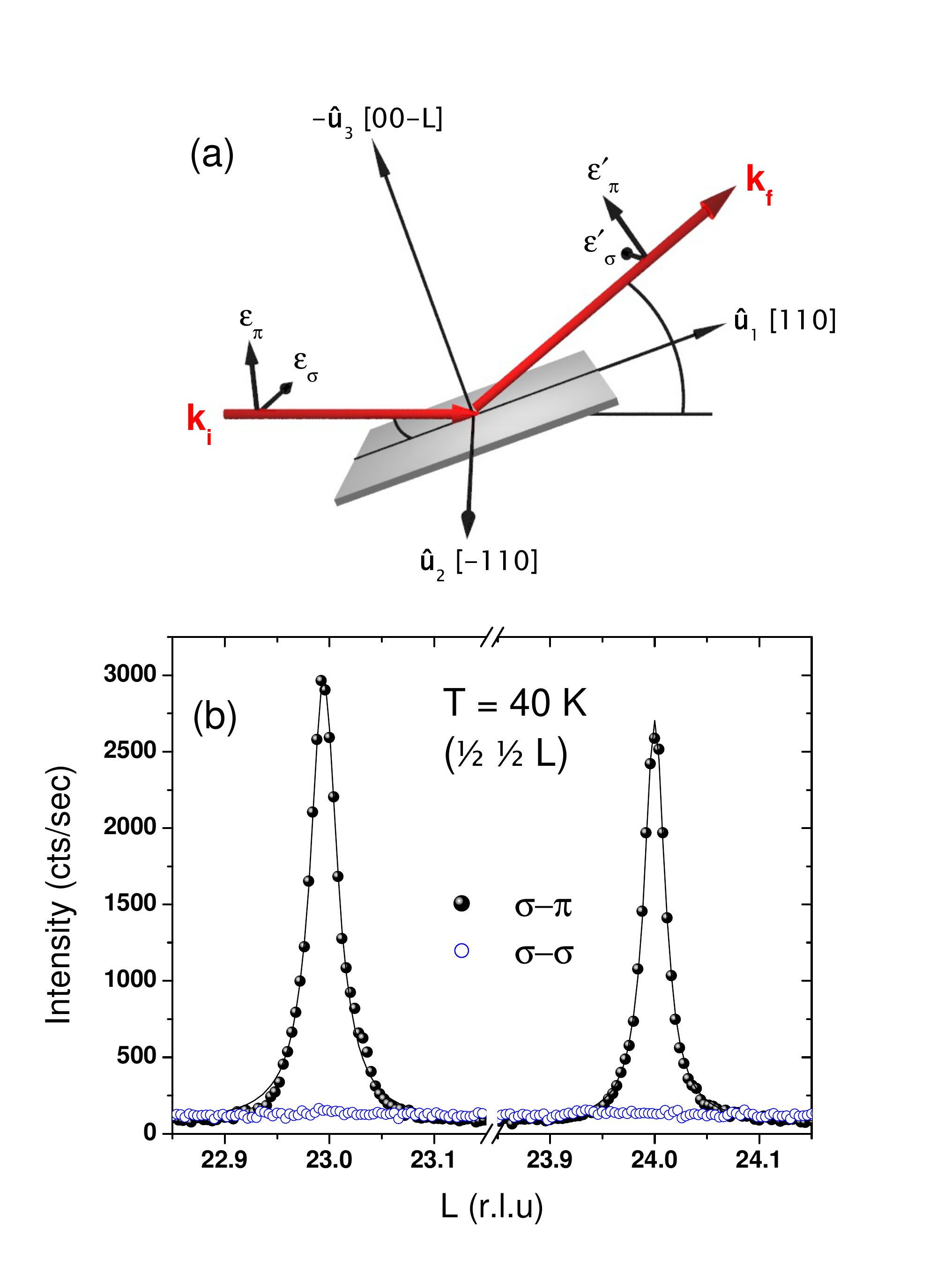}\caption
{\label{figure3}(Color online) (a) Schematic of the experimental configuration used for measuring XRS. The sample orientation is labelled with respect to the incoming and outgoing wavevectors. (b) L scan across the magnetic ($\mathrm{\frac{1}{2}\,\frac{1}{2}\,23}$) and ($\mathrm{\frac{1}{2}\,\frac{1}{2}\,24}$) reflections at the $\mathrm{L_{3}}$ edge, $\mathrm{T=40\,K}$, measured on I16. Both the $\mathrm{\sigma-\sigma}$ and $\mathrm{\sigma-\pi}$ polarization channels are shown; the blue open circles and black filled spheres, respectively. The solid line is a fit to a Lorentzian peak shape for each reflection in the $\mathrm{\sigma-\pi}$ channel. FWHM's of 0.031(4) and 0.024(3) r.l.u were found for the ($\mathrm{\frac{1}{2}\,\frac{1}{2}\,23}$) and ($\mathrm{\frac{1}{2}\,\frac{1}{2}\,24}$) reflections, respectively.}
\end{figure}

At the BM28 bending magnet beamline the energy was selected using a double-bounce Si (111) monochromator, and higher order harmonics were rejected by rhodium coated mirrors, providing a beam footprint of $\mathrm{\sim300\times800\mu m}$ at the sample. The diffractometer was a vertical scattering Eulerian cradle-type with a Vortex detector mounted on the $\mathrm{2\theta}$ (detector) arm. The degree of linear polarization in the plane of the storage ring (referred to as $\mathrm{\sigma}$-polarized light) is close to 100\% at I16 and BM28.

In both cases the samples were mounted in Displex cryostats with the [110] and [001] directions in the vertical scattering plane, the [001] being perpendicular to the sample surface. A schematic of the sample orientation can be seen in Fig. \ref{figure3} (a). A polarimeter was mounted on the detector arm; this consists of an analyzer crystal with a Bragg angle close to $\mathrm{45^{\circ}}$ at the selected energies and a suitable detector. By rotating the polarimeter set-up by $\mathrm{90^{\circ}}$ it is possible to detect $\mathrm{\pi}$-polarized light. In this way it is possible to discriminate between the two polarization channels of $\mathrm{\sigma-\pi}$ and $\mathrm{\sigma-\sigma}$ scattering, which means that different scattering mechanisms can be distinguished. For example, an incident x-ray beam of linearly polarized light is rotated by $\mathrm{90^{\circ}}$ when scattered by magnetic dipoles \cite{Hill}. A side-effect of using an analyzer to measure the polarization state of the scattered photons is that fluorescent photons, that are emitted isotropically several hundred eV below the absorption edge, are discriminated out, this may reduce overall signal strength, but dramatically improves the signal to noise ratio. For the Ir $\mathrm{L_{2,3}}$ edges a Au (333) reflection was used.

The crystal mosaic, determined from the full-width at half-maximum (FWHM) of the specular reflection ($\mathrm{0\,0\,24}$) in the unrotated $\mathrm{\sigma-\sigma}$ channel, was $\mathrm{0.044^{\circ}}$. In the $\mathrm{\sigma-\pi}$ rotated channel, peaks were found at the $\mathrm{(\frac{1}{2}\frac{1}{2}L)}$ positions. Examples of these peaks at the $\mathrm{(\frac{1}{2}\frac{1}{2}23)}$ and $\mathrm{(\frac{1}{2}\frac{1}{2}24)}$ reflections are shown in Fig. \ref{figure3}, as measured on I16 at 40\,K at the $\mathrm{L_{3}}$ edge, in reciprocal lattice units (r.l.u). $\mathrm{\sigma-\pi}$ intensity is shown as the full black circles and $\mathrm{\sigma-\sigma}$ as the open blue circles.

\subsubsection{Energy dependence and branching ratio}

The resonant enhancement of the ($\mathrm{\frac{1}{2}\,\frac{1}{2}\,24}$) reflection at $\mathrm{60\,K}$ was measured as a function of energy at the Ir $\mathrm{L_{2,3}}$ edges, from $\mathrm{\sim50\,eV}$ below the edge to $\mathrm{\sim50\,eV}$ above. The fluorescence was measured simultaneously in order to compare the energy dependence of the magnetic reflection intensity with the absorption white line.

\begin{figure}[htbp]
\centering
\includegraphics[width=0.475\textwidth,bb=50 10 790 530,clip]{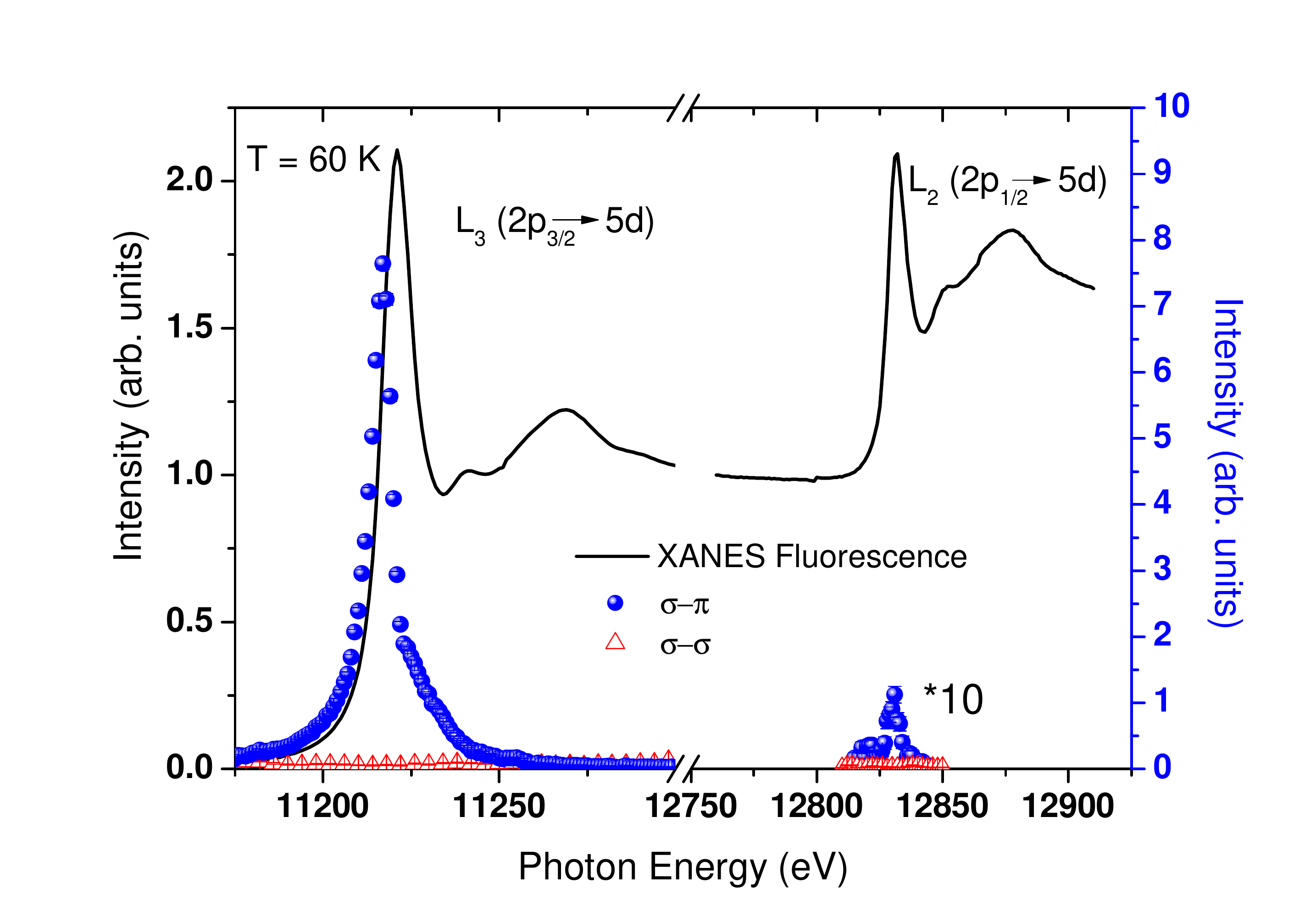}\caption
{\label{figure4}(Color online) Measurement of the resonant enhancement of the ($\mathrm{\frac{1}{2}\,\frac{1}{2}\,24}$) reflection across the $\mathrm{L_{2}}$ and $\mathrm{L_{3}}$ edges at $\mathrm{60\,K}$, well below $\mathrm{T_{N}}$. The solid black line shows the x-ray absorption near edge spectra (XANES), measured in fluorescence mode, normalized to the number of initial states. The blue solid points show the intensity of the ($\mathrm{\frac{1}{2}\,\frac{1}{2}\,24}$) peak in the rotated $\mathrm{\sigma-\pi}$ channel and the red open triangles result from the unrotated $\mathrm{\sigma-\sigma}$ intensity.}
\end{figure}

Figure \ref{figure4} shows the energy dependence of the ($\mathrm{\frac{1}{2}\,\frac{1}{2}\,24}$) magnetic reflection at the $\mathrm{L_{2,3}}$ edges. The XANES spectra (solid black line) has been included to reference the energy of the absorption edges. The contrast in intensity between the two edges is striking; similar to that reported for $\mathrm{Sr_{2}IrO_{4}}$ \cite{Kim2}, the integrated intensity of the magnetic scattering at the $\mathrm{L_{3}}$ edge is almost 30 times larger than that at the $\mathrm{L_{2}}$ (note that the signal at the $\mathrm{L_{2}}$ edge has been magnified by a factor of 10 for clarity). The exceptionally large ratio of the $\mathrm{L_{3}/L_{2}}$ edge resonant intensities is peculiar to the iridates thus far. Other \textit{5d} compounds, such as the conventional band insulator $\mathrm{K_{2}ReCl_{6}}$ \cite{McMorrow}, exhibit rather more modest ratios ($\mathrm{\sim2}$).

The widths of the $\mathrm{L_{2,3}}$ edge resonances are $\mathrm{FWHM_{L2}=5.29(5)\,eV}$ and $\mathrm{FWHM_{L3}=8.25(5)\,eV}$, respectively, similar to those found for $\mathrm{Sr_{2}IrO_{4}}$ \cite{Kim2} and other \textit{5d} materials \cite{McMorrow}. At the $\mathrm{L_{2}}$ edge the peak magnetic intensity is at the peak of the absorption edge white line, whereas at the $\mathrm{L_{3}}$ edge the resonance consists of two components, the most intense of which is centered on the inflection point of the absorption edge and the other at the peak of the white line. Energy scans at various temperatures about the ordering temperature indicated that both components share the same dipolar origin. On close inspection of the recent report on $\mathrm{Sr_{2}IrO_{4}}$ \cite{Kim2} by Kim \textit{et al.} the $\mathrm{L_{3}}$ edge resonance appears to present a similar two-component shape.

\subsubsection{Order parameter}

In order to determine the transition temperature of the magnetic ordering and the type of phase transition occurring in $\mathrm{Sr_{3}Ir_{2}O_{7}}$, we measured the intensity of the ($\mathrm{\frac{1}{2}\,\frac{1}{2}\,24}$) and ($\mathrm{\frac{1}{2}\,\frac{1}{2}\,23}$) reflections at the peak of the $\mathrm{L_{3}}$ edge resonance in both the unrotated $\mathrm{\sigma-\sigma}$ and the rotated $\mathrm{\sigma-\pi}$ channels.

\begin{figure}[h!]
\centering
\includegraphics[width=0.475\textwidth,bb=60 15 760 520,clip]{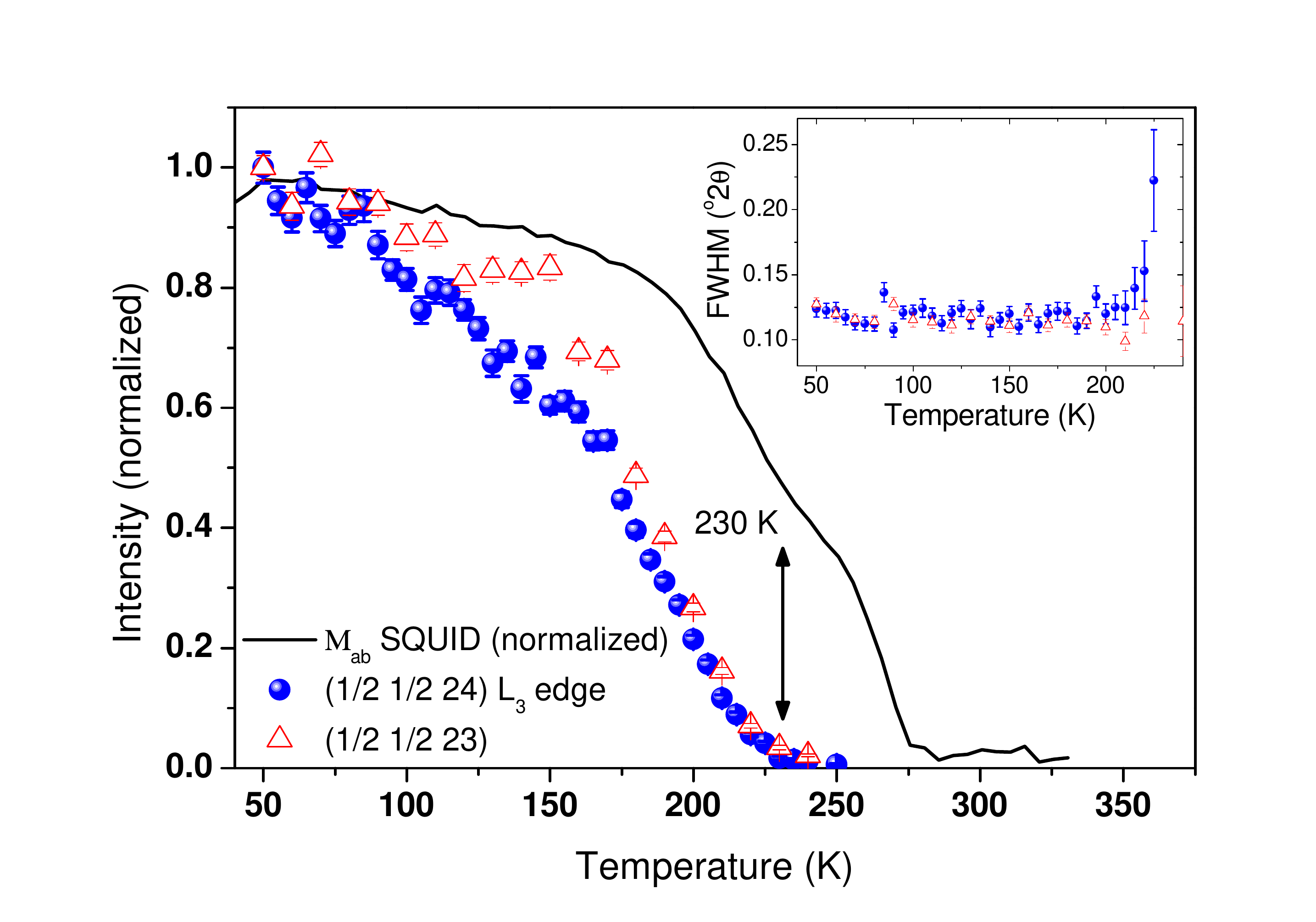}\caption
{\label{figure5}(Color online) The temperature dependence of the ($\mathrm{\frac{1}{2}\,\frac{1}{2}\,24}$) and ($\mathrm{\frac{1}{2}\,\frac{1}{2}\,23}$) peaks, measured at the $\mathrm{L_{3}}$ edge on the BM28 beamline is compared to the SQUID magnetization data. The transition temperature of the observed magnetic reflections, determined by XRS, coincides with the second transition, $\mathrm{T_{B}=230\,K}$ in the bulk magnetization data.}
\end{figure}

Fig. \ref{figure5} presents an integration of $\mathrm{\theta-2\theta}$ scans at the \textit{l}=23,24 peaks and is compared to
FC, SQUID magnetization data, measured in 0.005\,T. The transition appears second order with a N\'{e}el transition to a commensurate antiferromagnet, $\mathrm{T_{N}\approx230\,K}$. This is very close to the transition, $\mathrm{T_{B}}$, observed in the bulk magnetization data shown in Fig. \ref{figure5} and described in section \ref{magnetization}.

A question then remains as to the origin of the higher temperature transition, $\mathrm{T_{A}}$, observed in the bulk magnetization. It is possible that this indicates a transition to an incommensurate antiferromagnetic structure. Future measurements are planned to investigate the resonant magnetic scattering in this phase.

\subsubsection{Magnetic structure}

From the XRS investigation described previously, a series of peaks were found at $(\frac{h}{2}\frac{k}{2}l)$ positions, which result from a commensurate, ordered magnetic structure. Using the SARA\textit{h} computer program \cite{Wills}, we have used the approach of representational analysis, i.e. the application of Group Theory to solve the magnetic structure of $\mathrm{Sr_{3}Ir_{2}O_{7}}$.

\begin{figure}[h!]
\centering
\vbox{\subfigure{\includegraphics[width=0.25\textwidth,bb=70 90 400 660,clip]{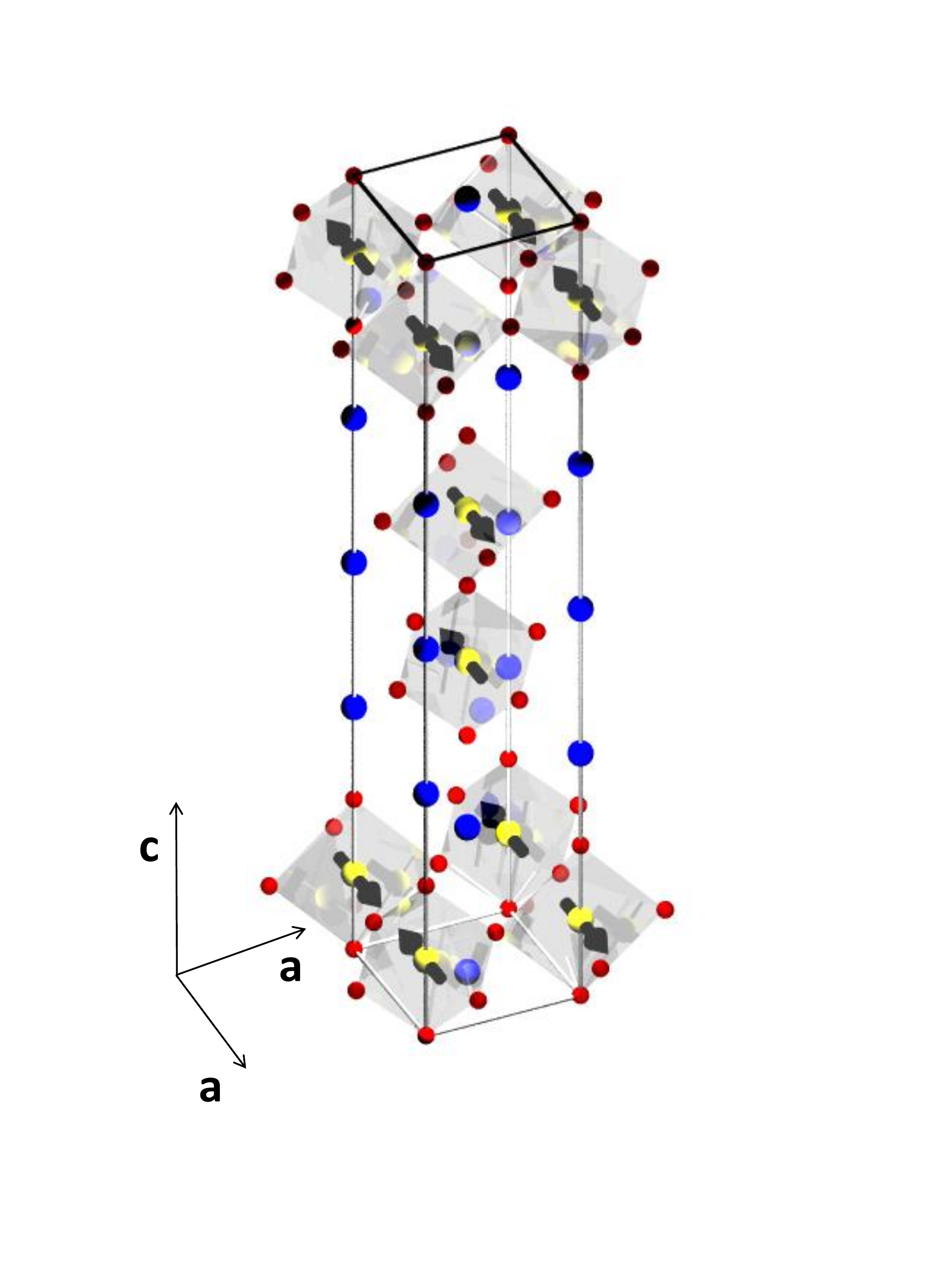}}\quad
\subfigure{\includegraphics[width=0.25\textwidth,bb=130 200 400 660,clip]{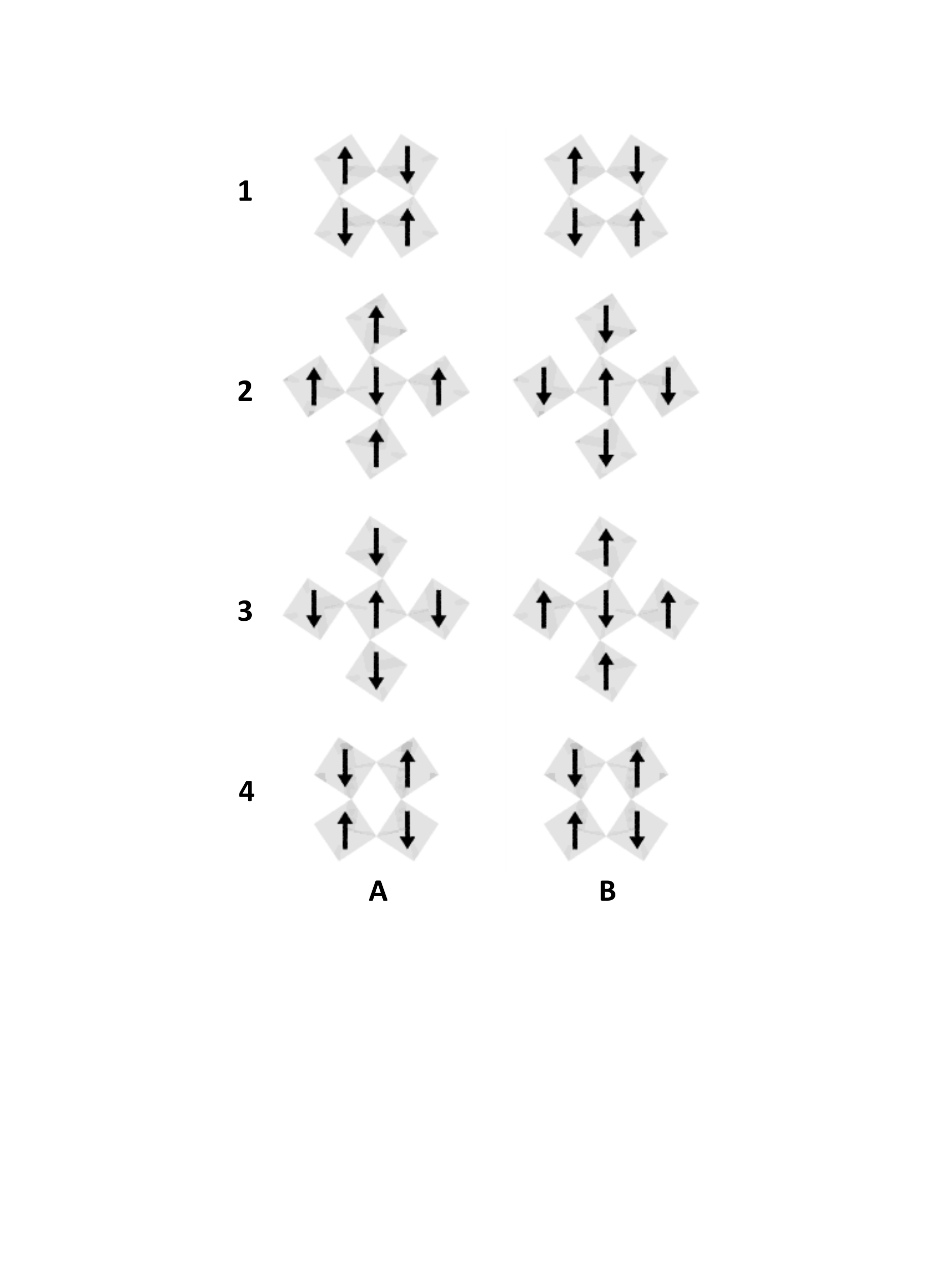}}}
\caption
{\label{figure6}(Color online) The upper panel shows the unit cell structure for the domain with ordering wavevector, $\mathrm{\mathbf{k_{1}}=(\frac{1}{2},\frac{1}{2},0)}$, as determined by Group Theory calculations performed using the program, SARA\textit{h} \cite{Wills} and the XRS data, presented in section \ref{xrs}. The lower panel shows a schematic representation of the two magnetic domains, A and B, defined by the ordering wavevectors, $\mathrm{\mathbf{k_{1}}=(\frac{1}{2},\frac{1}{2},0)}$ and $\mathrm{\mathbf{k_{2}}=(\frac{1}{2},-\frac{1}{2},0)}$.}
\end{figure}

In order to calculate the symmetry allowed relations between the moments, the only inputs required were the space group of the crystal structure, the atomic coordinates of the magnetic atoms and the propagation vector, \textbf{k}. The possible magnetic structures are shown in Fig. \ref{figure6} and are generated by two arms of the propagation vector $\mathrm{\mathbf{k}=(\frac{1}{2},\frac{1}{2},0)}$: $\mathrm{\mathbf{k_{1}}=(\frac{1}{2},\frac{1}{2},0)}$ and $\mathrm{\mathbf{k_{2}}=(\frac{1}{2},-\frac{1}{2},0)}$. The resulting magnetic structure may then consist of two domains, which have intensity for odd (even) \textit{l} peaks, respectively.

\subsubsection{Magnetic domain imaging}

In order to investigate the possibility that domains of odd (even) \textit{l} may exist in our samples we compared the intensities of several magnetic reflections. The similar intensities of the ($\mathrm{\frac{1}{2}\,\frac{1}{2}\,24}$) and ($\mathrm{\frac{1}{2}\,\frac{1}{2}\,23}$) peaks measured on the XMaS beamline, suggest that if the two-domain picture is correct then the domains are significantly smaller than the $\mathrm{300\times800\,\mu m}$ beam size, utilized on BM28. An approximately equivalent number of these domains would then be illuminated at each of the peak positions. On the I16 beamline however, the x-rays are focussed at the sample position. The beam spot is reduced, using a pair of parallel double focussing mirrors and slits, which results in a footprint on the sample surface of $\mathrm{(18\times100\,\mu m)}$. In this case we observed large differences in the intensities of odd (even) \textit{l} magnetic reflections.

In order to image the intensity of these magnetic reflections over a selected sample volume we performed a raster in x and y sample position at the \textit{l}=23,24 magnetic reflections, at the $\mathrm{L_{3}}$ edge peak resonance energy at $\mathrm{40\,K}$.

\begin{figure}[htbp]
\centering
\includegraphics[width=0.475\textwidth,bb=70 220 525 610,clip]{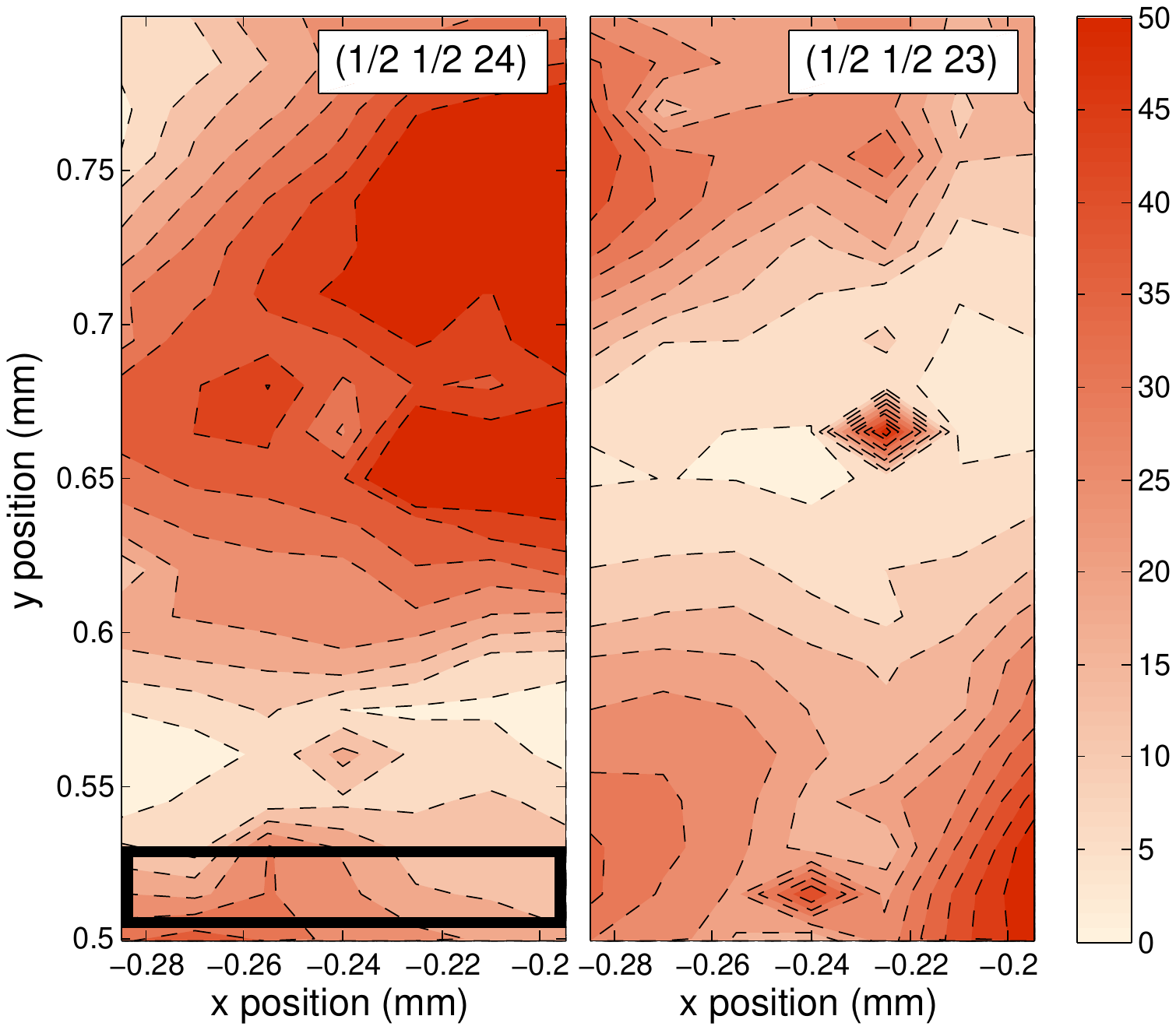}\caption
{\label{figure7}(Color online) Intensity of the ($\mathrm{\frac{1}{2}\,\frac{1}{2}\,24}$) reflection, left panel and ($\mathrm{\frac{1}{2}\,\frac{1}{2}\,23}$), right, as a function of x and y sample position. These measurements were made at the $\mathrm{L_{3}}$ edge resonance at a temperature of $\mathrm{40\,K}$, well below the N\'{e}el temperature, $\mathrm{T_{N}=230\,K}$. The beam size projected on the sample surface is highlighted by the black rectangle in the left-hand panel.}
\end{figure}

Figure \ref{figure7} presents contour plots of the ($\mathrm{\frac{1}{2}\,\frac{1}{2}\,24}$) and ($\mathrm{\frac{1}{2}\,\frac{1}{2}\,23}$) reflections over a $\mathrm{100\times300\,\mu m}$ area of the sample. The results are clear; in regions of the sample where there is little to no intensity of the \textit{l}=24 peak there is a maximum in the intensity of the \textit{l}=23 peak and vice versa. Domains of odd and even \textit{l} persist through the sample. This is not difficult to imagine, since the only difference between the two ordering wavevectors $\mathrm{\mathbf{k_{1}}}$ and $\mathrm{\mathbf{k_{2}}}$, is the respective orientation of one magnetic bilayer to another and these are of equivalent energy cost. These images indicate a domain size of approx. $\mathrm{100\times100\,\mu m}$ and confirm the calculations and results presented in section \ref{xrs}, supporting the determination of the magnetic structure.

\section{\label{FDMNES}Results of FDMNES calculations}

\begin{figure}[h!]
\centering
\includegraphics[width=0.475\textwidth,bb=70 20 550 530,clip]{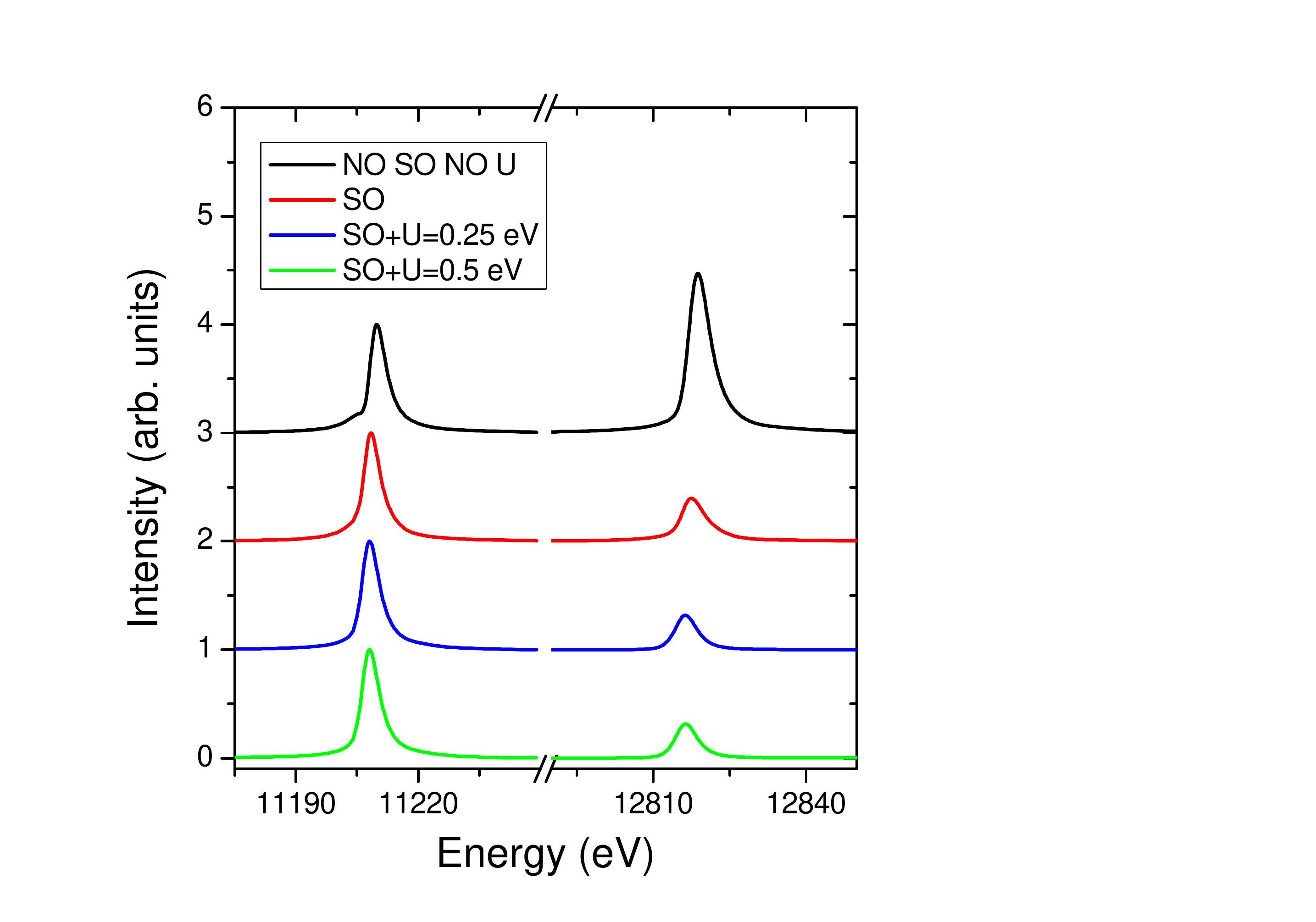}\caption
{\label{figure8}(Color online) FDMNES (finite difference method for solving the Schr\"{o}dinger equation) calculations performed at the Ir $\mathrm{L_{2,3}}$ edges for the $\mathrm{Sr_{3}Ir_{2}O_{7}}$ compound are shown for various initial parameters: assuming negligible SO and U (solid black line), including SO, but with no U (red line), including SO and U=0.25\,eV \cite{Moon} (blue line), and SO + U=0.5\,eV (green line). Spectra have been separated for clarity.}
\end{figure}

One of the most striking aspects of the x-ray resonant scattering investigations of both single and bilayered iridate compounds is the very large ratio of $\mathrm{L_{3}}$ and $\mathrm{L_{2}}$ edge magnetic scattering intensities. In order to better understand the origin of this remarkable effect we have performed calculations with the FDMNES code \cite{Joly}. FDMNES is an $\textit{ab initio}$ cluster-based, monoelectronic code that calculates x-ray absorption spectra and x-ray resonant scattering spectra.  We used the fully relativistic monoelectronic calculation (DFT-LSDA) with spin-orbit coupling on the basis of the Green's-function technique (multiple scattering) for a muffin-tin potential. In order to take account of the effect of the intersite Coulomb repulsion \textit{U}, not negligible in many theories of \textit{5d} Mott-insulators, we included the Hubbard correction (LSDA+\textit{U}). The value of U=0.25\,eV was chosen according to the optical conductivity measurements presented by Moon $\textit{et. al}$ \cite{Moon}. The simulated spectra derive entirely from the $E1-E1$ dipole interaction (we previously checked that there were no effects due to higher order terms). For the calculation we used a magnetic unit cell of size  $2a\times2b\times c$, containing 128 atoms with a cluster radius $\mathrm{3.8\,\AA}$ and an average of 19 atoms per cluster.

\begin{figure}[h!]
\centering
\includegraphics[width=0.475\textwidth,bb=30 140 760 1100,clip]{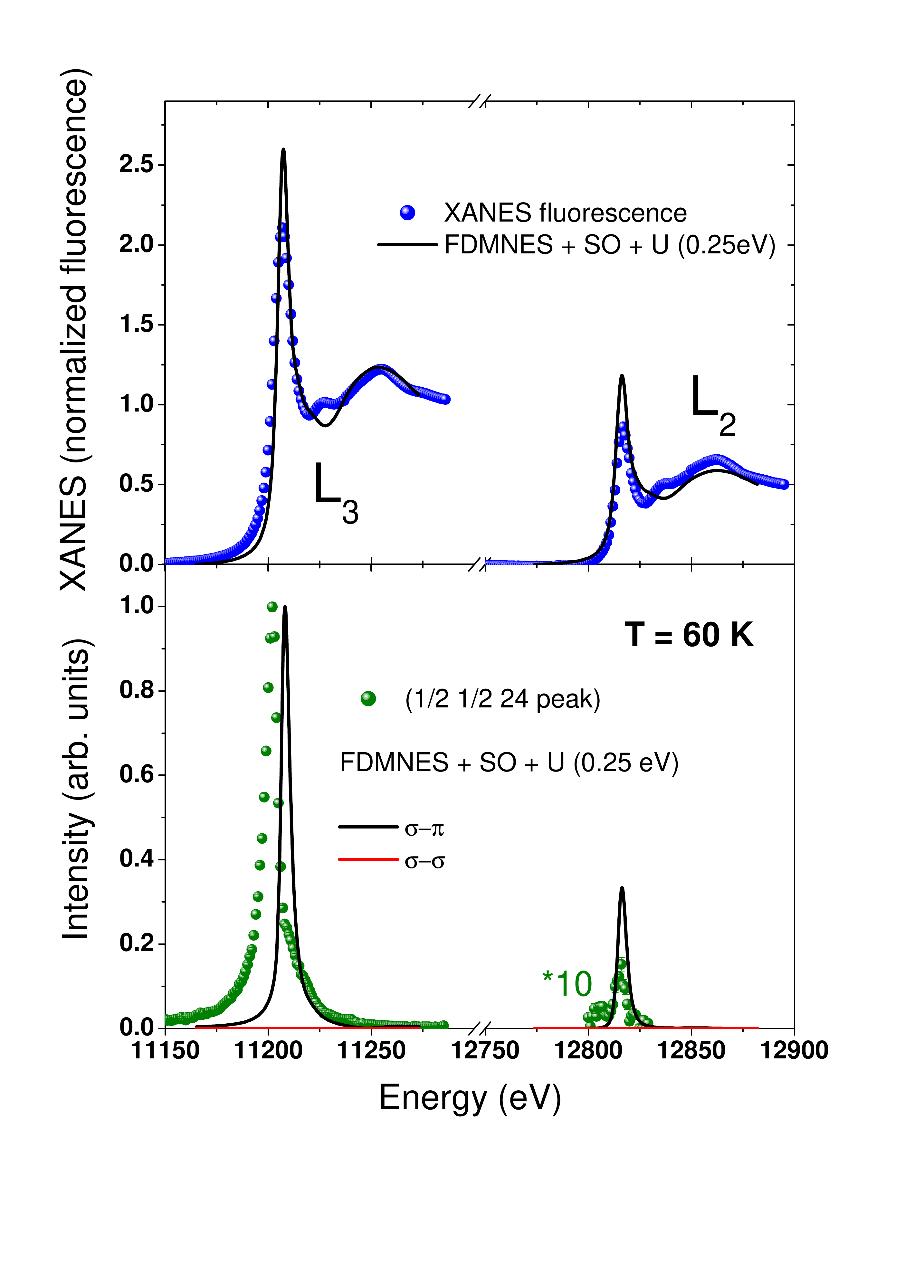}\caption
{\label{figure9}(Color online) FDMNES calculations for the XANES spectra and the ($\mathrm{\frac{1}{2}\,\frac{1}{2}\,24}$) magnetic reflection, including spin-orbit (SO) coupling and including both SO coupling and $\mathrm{U=0.25\,eV}$ are represented by solid black lines. The calculations are compared to the experimental XANES data at the $\mathrm{L_{2,3}}$ edges (blue solid points) and the ($\mathrm{\frac{1}{2}\,\frac{1}{2}\,24}$) magnetic reflection (green solid points).}
\end{figure}

Figure \ref{figure8} shows a comparison of several different calculations in order to observe the potential effects of the SO coupling and the Hubbard \textit{U} term on the relative intensities of the $\mathrm{L_{2,3}}$ edge resonances. It is clear that the spin-orbit coupling term plays the greatest role in the relative strengths of the $\mathrm{L_{3}}$ and $\mathrm{L_{2}}$ edge resonances. Including the Hubbard \textit{U} term dampens the $\mathrm{L_{2}}$ edge signal still further, but there is relatively little difference between 0.25\,eV and 0.5\,eV; the values calculated for $\mathrm{Sr_{3}Ir_{2}O_{7}}$ and $\mathrm{Sr_{2}IrO_{4}}$, respectively.

Figure \ref{figure9} compares the experimental data for the XANES spectra and x-ray resonances of magnetic reflections to FDMNES calculations. The upper panel presents the XANES data for $\mathrm{L_{3}}$ and $\mathrm{L_{2}}$ edges (blue solid points), whose edge jumps are normalized to the number of initial states, respectively. The FDMNES calculations solid black line) closely resemble the data; the intensity of the white line compared to the edge and the fine structure features in the XANES spectra are well-reproduced. The lower panel of Fig. \ref{figure9} shows the magnetic resonances of the $\mathrm{L_{2,3}}$ edges in the $\sigma-\pi$ channel, measured at $\mathrm{T=60\,K}$ at the ($\mathrm{\frac{1}{2}\,\frac{1}{2}\,24}$) magnetic reflection (solid green points). Some features are well modeled by the FDMNES and others are not; the ratio of $\mathrm{L_{3}}$ to $\mathrm{L_{2}}$ edge intensity is large ($\sim5$), but the damping of the signal at the $\mathrm{L_{2}}$ edge is not as extreme as that found in the experimental data. All of the scattered intensity appears in the $\sigma-\pi$ channel. However, the energy position of the maximum $\mathrm{L_{3}}$ edge resonance coincides with the peak in the white line absorption spectrum, not the inflection point, as is the case for the experimental data. These discrepancies may be due to the extreme spin-orbit coupling regime of $\mathrm{Sr_{3}Ir_{2}O_{7}}$ and the possible realization of a $\mathrm{J_{eff}=\frac{1}{2}}$ insulator.

\section{Discussion and conclusions}

This article presents a study of x-ray resonant scattering at the iridium $\mathrm{L_{2}}$ and $\mathrm{L_{3}}$ edges in bilayered $\mathrm{Sr_{3}Ir_{2}O_{7}}$. The principal aims were to investigate the magnetic structure and energy dependence of the magnetic reflections, in order to probe the robustness of the $\mathrm{L_{2}}$ edge damping observed in $\mathrm{Sr_{2}IrO_{4}}$, reported to be caused by a $\mathrm{J_{eff}=\frac{1}{2}}$ state \cite{Kim2}.

Calculations performed with the SARA\textit{h} \cite{Wills} programme, combined with the results from XRS data, indicate a two-domain commensurate antiferromagnetic (AF) structure with wavevectors $\mathrm{\mathbf{k_{1}}=(\frac{1}{2},\frac{1}{2},0)}$ and $\mathrm{\mathbf{k_{2}}=(\frac{1}{2},-\frac{1}{2},0)}$, although further azimuthal measurements are planned to determine the precise direction of the magnetic moments. The ordering temperature of this commensurate phase was found to coincide with the lower of two transitions, $\mathrm{T_{B}=230\,K}$, observed in the bulk magnetization data. The higher transition temperature may be due to an incommensurate AF phase.

The Lorenzian profile of the energy dependence is typical of a dipolar magnetic origin and only has intensity in the rotated, $\mathrm{\sigma-\pi}$ channel. FDMNES calculations, including spin-orbit coupling and the Hubbard term, \textit{U}, have reproduced many features of the XANES and XRS data, although the $\mathrm{L_{2}}$ edge damping is more extreme in the experimental data. The intensity of the $\mathrm{L_{3}}$ edge resonance is a factor of 30 times larger than the near negligible $\mathrm{L_{2}}$ edge, which mirrors results reported on the single layer cousin, $\mathrm{Sr_{2}IrO_{4}}$ \cite{Kim2}; an effect which is robust enough to overcome the increased bandwidth, \textit{W}, in the bilayered iridate. This could be explained by the $\mathrm{J_{eff}=\frac{1}{2}}$ model proposed by \textit{Kim et al.} \cite{Kim2} or that the observed moment is the \textit{5d} orbital angular momentum and not the total moment \cite{Chapon}.

\section{Acknowledgements}
We thank the Impact studentship programme, awarded jointly by UCL and Diamond Plc. for funding the thesis work of S. Boseggia. G. Nisbet and P. Thompson provided excellent instrument support at the I16 and BM28 beamlines, respectively. We also thank N. Casati for his support in the measurements and refinement process of the crystal structure determination. We thank A. S. Wills for useful discussions concerning the application of the SARA\textit{h} code to modelling the magnetic structure. We also acknowledge fruitful correspondence with Y. Joly, concerning the FDMNES calculations.

\bibliography{Sr3Ir2O7}

\begin{thebibliography}{26}
\expandafter\ifx\csname natexlab\endcsname\relax\def\natexlab#1{#1}\fi
\expandafter\ifx\csname bibnamefont\endcsname\relax
  \def\bibnamefont#1{#1}\fi
\expandafter\ifx\csname bibfnamefont\endcsname\relax
  \def\bibfnamefont#1{#1}\fi
\expandafter\ifx\csname citenamefont\endcsname\relax
  \def\citenamefont#1{#1}\fi
\expandafter\ifx\csname url\endcsname\relax
  \def\url#1{\texttt{#1}}\fi
\expandafter\ifx\csname urlprefix\endcsname\relax\def\urlprefix{URL }\fi
\providecommand{\bibinfo}[2]{#2}
\providecommand{\eprint}[2][]{\url{#2}}

\bibitem[{\citenamefont{Cheong}(2007)}]{Cheong}
\bibinfo{author}{\bibfnamefont{S.-W.} \bibnamefont{Cheong}},
  \bibinfo{journal}{Nat. Mat.} \textbf{\bibinfo{volume}{6}},
  \bibinfo{pages}{927} (\bibinfo{year}{2007}).

\bibitem[{\citenamefont{Bednorz and M\"{u}ller}(1986)}]{Bednorz}
\bibinfo{author}{\bibfnamefont{J.~G.} \bibnamefont{Bednorz}} \bibnamefont{and}
  \bibinfo{author}{\bibfnamefont{K.~A.} \bibnamefont{M\"{u}ller}},
  \bibinfo{journal}{Z. Phys. B - Condensed Matter}
  \textbf{\bibinfo{volume}{64}}, \bibinfo{pages}{189} (\bibinfo{year}{1986}).

\bibitem[{\citenamefont{Ramirez}(1997)}]{Ramirez}
\bibinfo{author}{\bibfnamefont{A.~P.} \bibnamefont{Ramirez}},
  \bibinfo{journal}{J. Phys.: Condens. Matter} \textbf{\bibinfo{volume}{9}},
  \bibinfo{pages}{8171} (\bibinfo{year}{1997}).

\bibitem[{\citenamefont{Cheong and Mostovoy}(2007)}]{Cheong2}
\bibinfo{author}{\bibfnamefont{S.-W.} \bibnamefont{Cheong}} \bibnamefont{and}
  \bibinfo{author}{\bibfnamefont{M.}~\bibnamefont{Mostovoy}},
  \bibinfo{journal}{Nat. Mat.} \textbf{\bibinfo{volume}{6}},
  \bibinfo{pages}{13} (\bibinfo{year}{2007}).

\bibitem[{\citenamefont{Kim et~al.}(2009)\citenamefont{Kim, Ohsumi, Komesu,
  Sakai, Morita, Takagi, and Arima}}]{Kim2}
\bibinfo{author}{\bibfnamefont{B.~J.} \bibnamefont{Kim}},
  \bibinfo{author}{\bibfnamefont{H.}~\bibnamefont{Ohsumi}},
  \bibinfo{author}{\bibfnamefont{T.}~\bibnamefont{Komesu}},
  \bibinfo{author}{\bibfnamefont{S.}~\bibnamefont{Sakai}},
  \bibinfo{author}{\bibfnamefont{T.}~\bibnamefont{Morita}},
  \bibinfo{author}{\bibfnamefont{H.}~\bibnamefont{Takagi}}, \bibnamefont{and}
  \bibinfo{author}{\bibfnamefont{T.}~\bibnamefont{Arima}},
  \bibinfo{journal}{Science} \textbf{\bibinfo{volume}{323}},
  \bibinfo{pages}{1329} (\bibinfo{year}{2009}).

\bibitem[{\citenamefont{Shitade et~al.}(2009)\citenamefont{Shitade, Katsura,
  Kune\v{s}, Qi, Zhang, and Hagaosa}}]{Shitade}
\bibinfo{author}{\bibfnamefont{A.}~\bibnamefont{Shitade}},
  \bibinfo{author}{\bibfnamefont{H.}~\bibnamefont{Katsura}},
  \bibinfo{author}{\bibfnamefont{J.}~\bibnamefont{Kune\v{s}}},
  \bibinfo{author}{\bibfnamefont{X.-L.} \bibnamefont{Qi}},
  \bibinfo{author}{\bibfnamefont{S.-C.} \bibnamefont{Zhang}}, \bibnamefont{and}
  \bibinfo{author}{\bibfnamefont{N.}~\bibnamefont{Hagaosa}},
  \bibinfo{journal}{Phys. Rev. Lett.} \textbf{\bibinfo{volume}{102}},
  \bibinfo{pages}{256403} (\bibinfo{year}{2009}).

\bibitem[{\citenamefont{Jiang et~al.}(2011)\citenamefont{Jiang, Gu, Qi, and
  Trebst}}]{Jiang}
\bibinfo{author}{\bibfnamefont{H.-C.} \bibnamefont{Jiang}},
  \bibinfo{author}{\bibfnamefont{Z.-C.} \bibnamefont{Gu}},
  \bibinfo{author}{\bibfnamefont{X.-L.} \bibnamefont{Qi}}, \bibnamefont{and}
  \bibinfo{author}{\bibfnamefont{S.}~\bibnamefont{Trebst}},
  \bibinfo{journal}{Phys. Rev. B} \textbf{\bibinfo{volume}{83}},
  \bibinfo{pages}{245104} (\bibinfo{year}{2011}).

\bibitem[{\citenamefont{Machida et~al.}(2010)\citenamefont{Machida, Nakatsuji,
  Onoda, Tayama, and Sakakibara}}]{Machida}
\bibinfo{author}{\bibfnamefont{Y.}~\bibnamefont{Machida}},
  \bibinfo{author}{\bibfnamefont{S.}~\bibnamefont{Nakatsuji}},
  \bibinfo{author}{\bibfnamefont{S.}~\bibnamefont{Onoda}},
  \bibinfo{author}{\bibfnamefont{T.}~\bibnamefont{Tayama}}, \bibnamefont{and}
  \bibinfo{author}{\bibfnamefont{T.}~\bibnamefont{Sakakibara}},
  \bibinfo{journal}{Nature} \textbf{\bibinfo{volume}{463}},
  \bibinfo{pages}{210} (\bibinfo{year}{2010}).

\bibitem[{\citenamefont{Crawford et~al.}(1994)\citenamefont{Crawford,
  Subramanian, Harlow, Fernandez-Baca, Wang, and Johnston}}]{Crawford}
\bibinfo{author}{\bibfnamefont{M.~K.} \bibnamefont{Crawford}},
  \bibinfo{author}{\bibfnamefont{M.~A.} \bibnamefont{Subramanian}},
  \bibinfo{author}{\bibfnamefont{R.~L.} \bibnamefont{Harlow}},
  \bibinfo{author}{\bibfnamefont{J.~A.} \bibnamefont{Fernandez-Baca}},
  \bibinfo{author}{\bibfnamefont{Z.~R.} \bibnamefont{Wang}}, \bibnamefont{and}
  \bibinfo{author}{\bibfnamefont{D.~C.} \bibnamefont{Johnston}},
  \bibinfo{journal}{Phys. Rev. B} \textbf{\bibinfo{volume}{49}},
  \bibinfo{pages}{9198} (\bibinfo{year}{1994}).

\bibitem[{\citenamefont{Cao et~al.}(2002)\citenamefont{Cao, Xin, Alexander,
  Crow, Schlottmann, Crawford, Harlow, and Marshall}}]{Cao}
\bibinfo{author}{\bibfnamefont{G.}~\bibnamefont{Cao}},
  \bibinfo{author}{\bibfnamefont{Y.}~\bibnamefont{Xin}},
  \bibinfo{author}{\bibfnamefont{C.~S.} \bibnamefont{Alexander}},
  \bibinfo{author}{\bibfnamefont{J.~E.} \bibnamefont{Crow}},
  \bibinfo{author}{\bibfnamefont{P.}~\bibnamefont{Schlottmann}},
  \bibinfo{author}{\bibfnamefont{K.}~\bibnamefont{Crawford}},
  \bibinfo{author}{\bibfnamefont{R.~L.} \bibnamefont{Harlow}},
  \bibnamefont{and} \bibinfo{author}{\bibfnamefont{W.}~\bibnamefont{Marshall}},
  \bibinfo{journal}{Phys. Rev. B} \textbf{\bibinfo{volume}{66}},
  \bibinfo{pages}{214412} (\bibinfo{year}{2002}).

\bibitem[{\citenamefont{Singh and Gegenwart}(2010)}]{Singh}
\bibinfo{author}{\bibfnamefont{Y.}~\bibnamefont{Singh}} \bibnamefont{and}
  \bibinfo{author}{\bibfnamefont{P.}~\bibnamefont{Gegenwart}},
  \bibinfo{journal}{Phys. Rev. B} \textbf{\bibinfo{volume}{82}},
  \bibinfo{pages}{064412} (\bibinfo{year}{2010}).

\bibitem[{\citenamefont{Kim et~al.}(2008)\citenamefont{Kim, Jin, Kim, Park,
  Leem, Yu, Noh, Kim, Oh, Park et~al.}}]{Kim}
\bibinfo{author}{\bibfnamefont{B.~J.} \bibnamefont{Kim}},
  \bibinfo{author}{\bibfnamefont{H.}~\bibnamefont{Jin}},
  \bibinfo{author}{\bibfnamefont{J.-Y.} \bibnamefont{Kim}},
  \bibinfo{author}{\bibfnamefont{B.-G.} \bibnamefont{Park}},
  \bibinfo{author}{\bibfnamefont{C.}~\bibnamefont{Leem}},
  \bibinfo{author}{\bibfnamefont{J.}~\bibnamefont{Yu}},
  \bibinfo{author}{\bibfnamefont{T.~W.} \bibnamefont{Noh}},
  \bibinfo{author}{\bibfnamefont{C.}~\bibnamefont{Kim}},
  \bibinfo{author}{\bibfnamefont{S.-J.} \bibnamefont{Oh}},
  \bibinfo{author}{\bibfnamefont{J.-H.} \bibnamefont{Park}},
  \bibnamefont{et~al.}, \bibinfo{journal}{Phys. Rev. Lett.}
  \textbf{\bibinfo{volume}{101}}, \bibinfo{pages}{076402}
  (\bibinfo{year}{2008}).

\bibitem[{\citenamefont{Nagai et~al.}(2007)\citenamefont{Nagai, Yoshida, Ikeda,
  Matsuhata, Kito, and Kosaka}}]{Nagai}
\bibinfo{author}{\bibfnamefont{I.}~\bibnamefont{Nagai}},
  \bibinfo{author}{\bibfnamefont{Y.}~\bibnamefont{Yoshida}},
  \bibinfo{author}{\bibfnamefont{S.-I.} \bibnamefont{Ikeda}},
  \bibinfo{author}{\bibfnamefont{H.}~\bibnamefont{Matsuhata}},
  \bibinfo{author}{\bibfnamefont{H.}~\bibnamefont{Kito}}, \bibnamefont{and}
  \bibinfo{author}{\bibfnamefont{M.}~\bibnamefont{Kosaka}},
  \bibinfo{journal}{J. Phys.: Condens. Matter} \textbf{\bibinfo{volume}{19}},
  \bibinfo{pages}{136214} (\bibinfo{year}{2007}).

\bibitem[{\citenamefont{Cao et~al.}(1998)\citenamefont{Cao, Bolivar, McCall,
  Crow, and Guertin}}]{Cao2}
\bibinfo{author}{\bibfnamefont{G.}~\bibnamefont{Cao}},
  \bibinfo{author}{\bibfnamefont{J.}~\bibnamefont{Bolivar}},
  \bibinfo{author}{\bibfnamefont{S.}~\bibnamefont{McCall}},
  \bibinfo{author}{\bibfnamefont{J.~E.} \bibnamefont{Crow}}, \bibnamefont{and}
  \bibinfo{author}{\bibfnamefont{R.~P.} \bibnamefont{Guertin}},
  \bibinfo{journal}{Phys. Rev. B} \textbf{\bibinfo{volume}{57}},
  \bibinfo{pages}{R11 039} (\bibinfo{year}{1998}).

\bibitem[{\citenamefont{Kini et~al.}(2006)\citenamefont{Kini, Strydom, Jeevan,
  Geibel, and Ramakrishnan}}]{Kini}
\bibinfo{author}{\bibfnamefont{N.~S.} \bibnamefont{Kini}},
  \bibinfo{author}{\bibfnamefont{A.~M.} \bibnamefont{Strydom}},
  \bibinfo{author}{\bibfnamefont{H.~S.} \bibnamefont{Jeevan}},
  \bibinfo{author}{\bibfnamefont{C.}~\bibnamefont{Geibel}}, \bibnamefont{and}
  \bibinfo{author}{\bibfnamefont{S.}~\bibnamefont{Ramakrishnan}},
  \bibinfo{journal}{J. Phys.: Condens. Matter} \textbf{\bibinfo{volume}{18}},
  \bibinfo{pages}{8205} (\bibinfo{year}{2006}).

\bibitem[{\citenamefont{Moon et~al.}(2008)\citenamefont{Moon, Jin, Kim, Choi,
  Lee, Yu, Cao, Sumi, Funakubo, Bernhard et~al.}}]{Moon}
\bibinfo{author}{\bibfnamefont{S.~J.} \bibnamefont{Moon}},
  \bibinfo{author}{\bibfnamefont{H.}~\bibnamefont{Jin}},
  \bibinfo{author}{\bibfnamefont{K.~W.} \bibnamefont{Kim}},
  \bibinfo{author}{\bibfnamefont{W.~S.} \bibnamefont{Choi}},
  \bibinfo{author}{\bibfnamefont{Y.~S.} \bibnamefont{Lee}},
  \bibinfo{author}{\bibfnamefont{J.}~\bibnamefont{Yu}},
  \bibinfo{author}{\bibfnamefont{G.}~\bibnamefont{Cao}},
  \bibinfo{author}{\bibfnamefont{A.}~\bibnamefont{Sumi}},
  \bibinfo{author}{\bibfnamefont{H.}~\bibnamefont{Funakubo}},
  \bibinfo{author}{\bibfnamefont{C.}~\bibnamefont{Bernhard}},
  \bibnamefont{et~al.}, \bibinfo{journal}{Phys. Rev. Lett.}
  \textbf{\bibinfo{volume}{101}}, \bibinfo{pages}{226402}
  (\bibinfo{year}{2008}).

\bibitem[{\citenamefont{Isaacs et~al.}(1990)\citenamefont{Isaacs, McWhan,
  Kleinman, Bishop, Ice, Zschack, Gaulin, Mason, Garrett, and Buyers}}]{Isaacs}
\bibinfo{author}{\bibfnamefont{E.~D.} \bibnamefont{Isaacs}},
  \bibinfo{author}{\bibfnamefont{D.~B.} \bibnamefont{McWhan}},
  \bibinfo{author}{\bibfnamefont{R.~N.} \bibnamefont{Kleinman}},
  \bibinfo{author}{\bibfnamefont{D.~J.} \bibnamefont{Bishop}},
  \bibinfo{author}{\bibfnamefont{G.~E.} \bibnamefont{Ice}},
  \bibinfo{author}{\bibfnamefont{P.}~\bibnamefont{Zschack}},
  \bibinfo{author}{\bibfnamefont{B.~D.} \bibnamefont{Gaulin}},
  \bibinfo{author}{\bibfnamefont{T.~E.} \bibnamefont{Mason}},
  \bibinfo{author}{\bibfnamefont{J.~D.} \bibnamefont{Garrett}},
  \bibnamefont{and} \bibinfo{author}{\bibfnamefont{W.~J.~L.}
  \bibnamefont{Buyers}}, \bibinfo{journal}{Phys. Rev. Lett.}
  \textbf{\bibinfo{volume}{65}}, \bibinfo{pages}{3185} (\bibinfo{year}{1990}).

\bibitem[{\citenamefont{Chapon and Lovesey}(2011)}]{Chapon}
\bibinfo{author}{\bibfnamefont{L.~C.} \bibnamefont{Chapon}} \bibnamefont{and}
  \bibinfo{author}{\bibfnamefont{S.~W.} \bibnamefont{Lovesey}},
  \bibinfo{journal}{J. Phys.: Condens. Matter} \textbf{\bibinfo{volume}{23}},
  \bibinfo{pages}{252201} (\bibinfo{year}{2011}).

\bibitem[{\citenamefont{Subramanian et~al.}(1994)\citenamefont{Subramanian,
  Crawford, and Harlow}}]{Subramanian}
\bibinfo{author}{\bibfnamefont{M.~A.} \bibnamefont{Subramanian}},
  \bibinfo{author}{\bibfnamefont{M.~K.} \bibnamefont{Crawford}},
  \bibnamefont{and} \bibinfo{author}{\bibfnamefont{R.~L.}
  \bibnamefont{Harlow}}, \bibinfo{journal}{Materials Research Bulletin}
  \textbf{\bibinfo{volume}{29}}, \bibinfo{pages}{645} (\bibinfo{year}{1994}).

\bibitem[{\citenamefont{Matsuhata et~al.}(2004)\citenamefont{Matsuhata, Nagai,
  Yoshida, Hara, Ikeda, and Shirakawa}}]{Matsuhata}
\bibinfo{author}{\bibfnamefont{H.}~\bibnamefont{Matsuhata}},
  \bibinfo{author}{\bibfnamefont{I.}~\bibnamefont{Nagai}},
  \bibinfo{author}{\bibfnamefont{Y.}~\bibnamefont{Yoshida}},
  \bibinfo{author}{\bibfnamefont{S.}~\bibnamefont{Hara}},
  \bibinfo{author}{\bibfnamefont{S.-I.} \bibnamefont{Ikeda}}, \bibnamefont{and}
  \bibinfo{author}{\bibfnamefont{N.}~\bibnamefont{Shirakawa}},
  \bibinfo{journal}{J. Solid State Chem.} \textbf{\bibinfo{volume}{177}},
  \bibinfo{pages}{3776} (\bibinfo{year}{2004}).

\bibitem[{\citenamefont{Blake et~al.}(2004)\citenamefont{Blake, Cao, and
  Radaelli}}]{Radaelli}
\bibinfo{author}{\bibfnamefont{G.~R.} \bibnamefont{Blake}},
  \bibinfo{author}{\bibfnamefont{G.}~\bibnamefont{Cao}}, \bibnamefont{and}
  \bibinfo{author}{\bibfnamefont{P.~G.} \bibnamefont{Radaelli}},
  \bibinfo{journal}{ESRF experimental report} pp. \bibinfo{pages}{HS--2386}
  (\bibinfo{year}{2004}).

\bibitem[{\citenamefont{Brown et~al.}(2001)\citenamefont{Brown, Bouchenoire,
  Bowyer, Kervin, Laundy, Longfield, Mannix, Paul, Stunault, Thompson
  et~al.}}]{Brown}
\bibinfo{author}{\bibfnamefont{S.~D.} \bibnamefont{Brown}},
  \bibinfo{author}{\bibfnamefont{L.}~\bibnamefont{Bouchenoire}},
  \bibinfo{author}{\bibfnamefont{D.}~\bibnamefont{Bowyer}},
  \bibinfo{author}{\bibfnamefont{J.}~\bibnamefont{Kervin}},
  \bibinfo{author}{\bibfnamefont{D.}~\bibnamefont{Laundy}},
  \bibinfo{author}{\bibfnamefont{M.~J.} \bibnamefont{Longfield}},
  \bibinfo{author}{\bibfnamefont{D.}~\bibnamefont{Mannix}},
  \bibinfo{author}{\bibfnamefont{D.~F.} \bibnamefont{Paul}},
  \bibinfo{author}{\bibfnamefont{A.}~\bibnamefont{Stunault}},
  \bibinfo{author}{\bibfnamefont{P.}~\bibnamefont{Thompson}},
  \bibnamefont{et~al.}, \bibinfo{journal}{J. Synchrot. Radiat.}
  \textbf{\bibinfo{volume}{8}}, \bibinfo{pages}{1172} (\bibinfo{year}{2001}).

\bibitem[{\citenamefont{Hill and McMorrow}(1996)}]{Hill}
\bibinfo{author}{\bibfnamefont{J.~P.} \bibnamefont{Hill}} \bibnamefont{and}
  \bibinfo{author}{\bibfnamefont{D.~F.} \bibnamefont{McMorrow}},
  \bibinfo{journal}{Acta Crystallogr.} \textbf{\bibinfo{volume}{52}},
  \bibinfo{pages}{236} (\bibinfo{year}{1996}).

\bibitem[{\citenamefont{McMorrow et~al.}(2003)\citenamefont{McMorrow, Nagler,
  McEwen, and Brown}}]{McMorrow}
\bibinfo{author}{\bibfnamefont{D.~F.} \bibnamefont{McMorrow}},
  \bibinfo{author}{\bibfnamefont{S.~E.} \bibnamefont{Nagler}},
  \bibinfo{author}{\bibfnamefont{K.~A.} \bibnamefont{McEwen}},
  \bibnamefont{and} \bibinfo{author}{\bibfnamefont{S.~D.} \bibnamefont{Brown}},
  \bibinfo{journal}{J. Phys.: Condens. Matter} \textbf{\bibinfo{volume}{15}},
  \bibinfo{pages}{L59} (\bibinfo{year}{2003}).

\bibitem[{\citenamefont{Wills}(2000)}]{Wills}
\bibinfo{author}{\bibfnamefont{A.~S.} \bibnamefont{Wills}},
  \bibinfo{journal}{Physica B} \textbf{\bibinfo{volume}{680}},
  \bibinfo{pages}{276} (\bibinfo{year}{2000}).

\bibitem[{\citenamefont{Joly}(2001)}]{Joly}
\bibinfo{author}{\bibfnamefont{Y.}~\bibnamefont{Joly}}, \bibinfo{journal}{Phys.
  Rev. B} \textbf{\bibinfo{volume}{63}}, \bibinfo{pages}{125120}
  (\bibinfo{year}{2001}).

\end{thebibliography}

\end{document}